\documentclass[12pt]{article}
\usepackage{latexsym}
\usepackage{graphics}
\usepackage{epsfig}
\usepackage{epstopdf}
\usepackage{amsmath}
\usepackage{cite}
\usepackage{hyperref}
\usepackage{amssymb}

\begin{document}

\begin{center}
\textbf{\Large 
Influence of relaxation on filtering microflows under harmonic action on the layer 
} \vspace{0.5 cm}

 I.I. Denysiuk$^1$,  I.A. Skurativska$^1$\footnote{e-mail:
\url{inna.skurativska@gmail.com}},  I.V. Bielinskyi$^1$,   O.M. Sizonenko$^2$, 
   I.M. Hubar$^1$  \vspace{0.5 cm}

$^1$ S.I.Subbotin Institute of Geophysics of the National Academy of Science of Ukraine, Kyiv, Ukraine
     
$^2$     Institute of Impulse Processes and Technologies of the National Academy of Science of Ukraine, Mykolaiv, Ukraine 
\end{center}

\begin{quote} \textbf{Abstract.}{\small 

{\bf Purpose.} Investigation of the velocity fields of non-equilibrium fluid filtration in a layer under harmonic action on it and assessment of the influence of relaxation effects on the attenuation of the amplitude of the initial disturbance within the framework of mathematical modeling of non-equilibrium plane-radial filtration.

{\bf Methodology.} A mathematical model of non-equilibrium plane-radial filtration with a generalized dynamic Darcy law in the form of a boundary value problem in a half-space with a harmonic excitation law at its boundary is considered. Based on the exact solutions of the boundary value problem, the attenuation of the amplitude of the initial disturbance from the model's parameters and their influence on the size of the disturbed area is investigated.

{\bf Findings.} A differential equation modeling non-equilibrium filtration processes in the massif in the cylindrical reference frame was obtained. Using the method of separation of variables, it was constructed the solution, bounded at infinity, to the model differential equation subjected to harmonic action at the layer boundary. The solution's asymptotic approximation was constructed for large values of the argument. Using the asymptotic solution of the boundary value problem, the damping of velocity field pulsations during non-equilibrium filtration was analyzed depending on the frequency of the harmonic action, the ratio of the piezoconductivity coefficients of the layer and the relaxation time. Profiles of the dependences of the size of the influence zone on the model parameters were plotted and the choice of parameters for optimal influence on the bottom-hole zone of the well was analyzed.

{\bf Originality.} On the basis of the non-equilibrium filtration model, it is shown that harmonic disturbances applied to the boundary of a semi-infinite layer can penetrate the reservoir over a greater distance under the conditions of manifestation of the relaxation mechanism of the fluid-skeleton interaction, compared to the equilibrium filtration process. Such an effect is observed at a finite interval of disturbance frequencies, while at high frequencies relaxation contributes to a more significant damping of disturbances. In the parametric space of excitation frequency - relaxation time, there is a locus of points that corresponds to the maximum sizes of influence zone of disturbances.

{\bf Practical value.} The obtained results are relevant for research on the impact of wave disturbances on the layer with the aim of intensifying filtration processes, as well as for the creation of new wave technologies to increase the extraction of mineral resources from productive layers.
}
\end{quote}

\begin{quote} \textbf{Keyword:}{\small 
 non-equilibrium filtration, Darcy's generalized law, porous medium, wave action, attenuation, filtration velocity fields.
}
\end{quote}

\vspace{0.5 cm}

\section*{Introduction}

	 At the present, the energy stability of Ukraine is connected with the increase in the production of energy both due to the development of new promising hydrocarbon deposits and the use of new technologies for the enhancing their production.
In the process of developing oil and gas fields, the filtration characteristics of the rocks serving as oil and gas collectors deteriorate significantly, which leads to a decrease in the flow rate of wells and the degree of mineral resources development.

Most of the highly productive fields of the oil and gas complex of Ukraine have entered the final stage of development, which is characterized by the progressive depletion of formation energy, the watering of wells and the increase in the share of hard-to-extract reserves. The development of fields with difficult-to-extract oil reserves is carried out at a low rate, and the final yield of oil in such cases does not exceed 30-40\% of the initial balance of their reserves \cite{ref01}. 

In this regard, the most important scientific and technical problem that arises during the exploitation of deposits is the most complete extraction of oil while ensuring high rates of development. Therefore, the tasks of applying new oil production technologies that allow to significantly increase the oil yield of the layers being developed, from which it is no longer possible to extract significant residual oil reserves using traditional methods, are urgent.

To extract these remaining oil reserves, various modern methods of intensification of oil production are used, in particular, thermal, chemical, physical, biological and others. Among these methods, physical methods of increasing the flow rate of production wells play a leading role. They also include methods of wave action on the near wellbore zone (NWZ) and on the formation as a whole. Methods of wave action can be divided into several groups: acoustic (ultrasonic, hydraulic), shock wave and vibroseismic ones  \cite{ref02}.

Knowledge of physical processes and phenomena that are responsible for restoring the filtration properties of reservoir rocks and fluid mobility is necessary for a well-founded choice of the method of wave action on the reservoir and increasing its potential oil extraction capabilities. The rate of filtration, which is determined using methods of mathematical modeling of filtration processes, is used to quantify fluid mobility.
Therefore, the study of the mechanisms of the influence of the wave field to restore the filtration properties of the formation, to solve the problems of increasing oil recovery is an urgent task. The advantages of the methods of wave action on the formation are considered to be, firstly, the possibility of adjusting the parameters of the emitter of acoustic waves (intensity, frequency, duration of treatment); secondly, ecological purity of the method; thirdly, its high efficiency.

Literature review. To study the mechanisms of wave action on saturated porous reservoir media and reservoir fluids, as well as to identify the peculiarities of the propagation of elastic waves in heterogeneous media, many experimental studies have been performed. In particular, in works  \cite{ref03,ref04}, the effect of ultrasonic waves on the rheological characteristics of various samples of oil and its processing products was experimentally investigated. Experiments proved that the viscosity of all liquid samples decreased during the action of ultrasonic disturbances. It was found  \cite{ref03} that a higher power of ultrasound leads to a more intense decrease in the viscosity of the liquid.

The results of laboratory studies  \cite{ref04} also showed that the mobility of fluids increases under ultrasonic action, while the authors of the studies \cite{ref05}, in addition to a decrease in fluid viscosity, also observed an increase in reservoir permeability. The effect of increasing the absolute permeability of saturated porous reservoir media under the influence of high-amplitude pressure fluctuations in the liquid was experimentally proven in \cite{ref06}.

The publication \cite{ref07} provides experimental studies on model heterogeneous media with the aim of confirming the increase in oil mobility under ultrasonic action in porous media. According to the results of these studies, it was established that the oil extraction coefficient is proportional to the power of the wave emitter and depends on the frequency of ultrasonic action.

In order to achieve the maximum efficiency of the wave action on the formation and NWZ, it is also advisable to use pulse-wave methods with the formation of shock pulses in a fluid-filled well, starting from the wellhead, using special devices - pulse generators. In \cite{ref08}, it is proposed to implement pulse-wave action using a hydraulic generator with a repetition frequency of pressure pulses 1-50 Hz, which are less intensively absorbed by the reservoir at distances exceeding 2 m from the well. As a result of the pulse-wave action on the productive layer, an optimal level of depression is formed, which contributes to the process of oil extraction and purification of NWZ. To do this, special pumping equipment is used, which allows one to smoothly change the pressure on the wellbore over a wide range. As a result, the joint use of a pump and a hydraulic pulse generator ensures the minimum content of clogging substances and the optimal mode of extracting reservoir fluids \cite{ref08}. For the purification of NWZ from colmatant, work \cite{ref09} proposes a method of shock-wave treatment, which is effectively used in the fields of Kazakhstan. The basis of the method is the use of a special device that forms rarefaction waves. The use of the shock-wave treatment method together with chemical compounds showed high efficiency in the process of cleaning the near-breakout zone of the formation from clogging substances and increasing the acceptability of injection wells several times (from 20 to 160 m$^3$/day).

In order to study the mechanism of interaction of elastic waves with the porous medium of the formation, the effect of elastic oscillations on the change of fluid filtration in the bulk model for the formation is considered in \cite{ref10}. The results of experimental studies indicate a significant influence of the field of elastic oscillations on the filtration of the oil-water mixture. In addition, during the experiments, other effects were observed: a decrease in the effective friction between the rock and the fluid, and as a result, an increase in the mobility of the fluid. During the passage of an elastic wave, stretching and compression phases are observed, which together with vibration of the skeleton and liquid causes the effect of a vibration pump; a change in the shape of the meniscus at the boundary of the separation of two phases and, as a result, a change in capillary pressure \cite{ref10}.

Understanding the wave process patterns occurring in the well-reservoir system is incomplete without theoretical research. Such studies are based on the analysis of the equations of the continuum mechanics for models of the real medium where a wave process is observed.

The authors of \cite{ref11} developed a mathematical modeling of the method of acoustic stimulation of wells. The model takes into account the following physical processes: reduction of liquid viscosity due to mixing and heating; excitation of elastic waves on the walls of the well (to reduce the adhesion forces between formation fluids and rock); excitation of natural frequencies associated with the vibration of the liquid inside the porous medium. Using numerical modeling, the optimal radiation frequencies were determined. It is shown that the well's productivity can be significantly improved due to the correct selection of operating frequencies of the acoustic emitter.
The publication \cite{ref12} deals with the problem of pulse-wave influence on a branched horizontal well, which is modeled by branched waveguides of a certain radius, in each of which the fluid movement is described by a wave equation. It is shown that there are resonant frequencies in branched wells, at which the pressure value can exceed the amplitude of the applied pressure pulse by several orders of magnitude. It was established that after pulse-wave treatment of injection wells in the fields of Oman, their acceptability increased almost three times.

On the basis of a model differential equation that takes into account the oscillation damping, the propagation of ultrasonic waves in a viscous liquid was studied in \cite{ref13} using the integral Laplace transform. Mathematical modeling of ultrasonic wave propagation made it possible to highlight the mechanisms of wave absorption and to study the influence of various parameters (temperature, relaxation time) on the propagation of waves in viscous liquids. The results of theoretical studies were verified experimentally by studying the change in the velocity of ultrasound propagation and the damping parameter in glycerol depending on temperature and frequency.

The results of research on the pulsating steady motions of a viscous fluid in the pore channels of the reservoir under the harmonic action of acoustic waves on them are presented in \cite{ref14}. To solve the problem, differential equations describing the laminar motion of a viscous liquid in a cylindrical pore channel were used. On the basis of the obtained solution, numerical calculations of dynamic processes in the pore channels of the formation were carried out. It was found that in the case of acoustic action on the formation, the speed of fluid movement in pore channels reaches maximum values in a certain frequency range, depending on the size of the pores and the kinematic viscosity of the fluid.

The article \cite{ref15} is devoted to the pulsating movements of a viscous fluid in the filtration channels of capillary-porous bodies under the action of harmonic waves. Pulsating movements are accompanied by compression-discharge waves and sign-changing filtration flows in the filtration channels of capillary-porous bodies. According to the results of theoretical studies, the most effective mode of pulse-wave loading was selected depending on the radius of the pore channel.

It should be noted that the researches on this topic known in the literature were conducted mainly on the basis of the classical Darcy filtration law without taking relaxation processes into account. However, to date, a considerable amount of experimental evidence of deviations from Darcy's linear law has been collected \cite{ref16}, especially in relation to non-equilibrium high-intensity processes, when the strengthening of non-local effects is observed \cite{ref17}.

To eliminate this gap, the article \cite{ref18} proposed a mathematical model for the elastic mode of liquid filtration with a generalized dynamic Darcy filtration law, which includes a description of nonlocal and nonlinear effects. Within the framework of this model, the influence of relaxation and the ratio of permeability coefficients of the reservoir rock on the phase speed of propagation of small wave disturbances was analyzed.
Thus, on the basis of the analysis of recent publications and research on this problem, it was established that a significant part of them is devoted to the elucidation of physical mechanisms and phenomena that affect the restoration of the filtration characteristics of the porous medium of the reservoir, the fluid rheological parameters and the increase of fluid mobility under wave action.

At the same time, the survey of scientific information proved that the study of the attenuation of filtration oscillations, which are formed in the process of acoustic action and the study of the influence of relaxation effects on the processes of non-equilibrium filtration, are at the initial stage of their study and are insufficiently covered. And these factors play a significant role in the development of methods of wave action on the porous medium of the formation in order to intensify the filtration processes in them. In this regard, the purpose of the research is to study on the basis of Darcy's generalized dynamic filtration law of pulsating damping filtration microflows of a fluid in a porous semi-bounded medium of an oil-bearing formation under harmonic action on it and to determine the critical frequency of the wave action, which ensures the minimum of the reduced attenuation coefficient depending on the ratio permeability coefficients and the relaxation parameter.

\section{The problem statement}
 It is considered the porous medium with specified initial porosity $m_0$,pores of which are filled with fluid. Let the emitter of acoustic harmonic waves acting on the formation be located in the well of the radius $r_{c}$ at the formation level. Then the fluid elements, located at a distance $r>r_{c}$  from the source of oscillations, perform a one-dimensional axisymmetric radial non-stationary periodic motion according to the law determined by the solution of the non-equilibrium filtration equation, which can be deduced \cite{ref18, refJPS} from the following system of equations written in cylindrical coordinates:
\begin{equation}\label{ds:eq1}
\begin{split}
 \tau\left[u_{t}+u \cdot u_{r}+\frac{k_{f}}{\mu}\left(p_{r t}+u p_{r r}\right)\right]+u+\frac{k_{e}}{\mu} p_{r}=0,\\
(m \rho)_{t}+u \rho_{r}+\rho\left(u_{r}+\frac{u}{r}\right)=0,\\
\rho=\rho_{0}+\rho_{0} \beta_{0}\left(p-p_{0}\right), \qquad
m=m_{0}+\beta_{s}\left(p-p_{0}\right), \\
\mu(p, T)= const, \quad k(p, r)= const,
\end{split}
\end{equation}
where  $u$ is the filtration velocity, m/s; $p$, $p_{0}$  are the variable and initial pressures respectively, Pa; $\rho$, $\rho_{0}$  are the variable and initial fluid densities, kg/m$^3$; $m$, $m_{0}$  are the variable and initial rock porosities; $\beta_{0}$,  $\beta_s$ are coefficients of volume compressibility of oil and rock skeleton of formation, 1/Pa; $\mu$   is the coefficient of dynamical viscosity of oil, Pa$\cdot$s;  $k_{e}$, $k_{f}$  are the steady and frozen coefficients of permeability, m$^2$;  $\tau$ is the time of relaxation, s.

\section{Description of the research methodology.}

  System of equations (\ref{ds:eq1}) is the essentially nonlinear and its analytical solutions are unknown till now. However, for small deviations from the equilibrium state that occurs under acoustic wave action on the reservoir, filtration processes with relaxation can be described with sufficient accuracy using the proper linearized equations.

Thus, using approximate methods of perturbation theory and methods of mathematical analysis, the nonlinear system of equations (\ref{ds:eq1}) is reduced to the following form
\begin{equation}\label{ds:eq2}
\begin{split}
\tau\left(u_{t}+\frac{k_{f}}{\mu} p_{r t}\right)+u+\frac{k_{e}}{\mu} p_{r}=0,  \\
\beta p_{t}+u_{r}+\frac{u}{r}=0,
\end{split}
\end{equation}
where  $\beta=m_{0} \beta_{0}+\beta_{s}$.

Eliminating the variable   from equations (\ref{ds:eq2}), we obtain the model equation describing non-equilibrium filtration processes of a fluid during plane radial motion
\begin{equation}\label{ds:eq3}
\begin{split}
 \tau K_{f} u_{r r t}+K_{e} u_{r r}+\tau K_{f} \frac{1}{r} u_{r t}+K_{e} \frac{u_{r}}{r}-K_{e} \frac{u}{r^{2}}- \tau u_{t t}-u_{t}\left(1+\tau K_{f} \frac{1}{r^{2}}\right)=0, 
\end{split}
\end{equation}
where $K_{e}=\frac{k_{e}}{\beta \mu}, \mathrm{m} / \mathrm{s}^{2}$;  $K_{f}=\frac{k_{f}}{\beta \mu}, \mathrm{m} / \mathrm{s}^{2}$  are the coefficients of piezoconductivity of formation in equilibrium and frozen states respectively.

Presentation of research results and discussion. To study the effects of wave harmonic action on filtration in the porous medium of an oil-bearing reservoir, it is necessary to determine the solution of equation (\ref{ds:eq3}) with the following boundary condition
\begin{equation}\label{ds:eq4}
u(r, t)=A \sin \omega t \text { at } r=r_{c}.
\end{equation}
The constrain providing boundedness of equation's solution at infinity reads as follows 
$$
\lim _{r \rightarrow \infty} u(r, t)=0 .
$$
It is convenient to look for the solution of the boundary value problem in the class of complex-valued functions (with the appropriate modification of the boundary condition) by the Fourier method (method of separation of variables) in the form:
\begin{equation}\label{ds:eq5}
\tilde{u}(r, t)=R(r) \cdot e^{i \omega t}, 
\end{equation}
where  $\omega$ is the circular frequency, $\mathrm{s}^{-1}$.

Inserting (\ref{ds:eq5}) into equation (\ref{ds:eq3}), we obtain the Bessel equation with complex argument 
\begin{equation}\label{ds:eq6}
\frac{d^{2} R}{d r^{2}}+\frac{1}{r} \frac{d R}{d r}-R\left(a^{2}+\frac{1}{r^{2}}\right)=0,
\end{equation}
where  $a^{2}=\frac{\omega}{\mathrm{K}_{e}}\left(\frac{i-\omega \tau}{1+i \omega \tau \theta}\right)$ and $\theta=K_{f} / K_{e}$.

In general, the parameter  $\theta$ can be regarded as a measure of how far relaxation can take the system from the equilibrium state or how much the systems can differ in the equilibrium and frozen states. Let us recall that $\theta<1$  according to thermodynamic constraints \cite{refKoster}.

Thus, the solution of equation (\ref{ds:eq6}) with modified boundary condition $R(r)=A$ at $r=r_{c}$ and condition at infinity $R(r) \rightarrow 0$ has the form
\begin{equation}\label{ds:eq7}
R(r)=A \frac{\mathrm{K}_{1}(r a)}{\mathrm{K}_{1}\left(r_{c} a\right)}, 
\end{equation}
where $\mathrm{K}_{1}(\cdot)$ is the modified Bessel function of the second kind of the first order.

The number  $a$ in (\ref{ds:eq7}) is defined by the well-known relation $a=\varphi+i \alpha$, where
\begin{align}
& \varphi=\sqrt{\frac{\omega}{2 \mathrm{~K}_{e}}} \sqrt{\sqrt{\frac{1+z^{2}}{1+\theta^{2} z^{2}}}+\frac{z(\theta-1)}{1+\theta^{2} z^{2}}}>0, \label{ds:eq8}\\
& \alpha=\sqrt{\frac{\omega}{2 \mathrm{~K}_{e}}} \sqrt{\frac{1+z^{2}}{1+\theta^{2} z^{2}}-\frac{z(\theta-1)}{1+\theta^{2} z^{2}}}, \label{ds:eq9}
\end{align}
%
and $z=\omega \tau$ is the dimensionless frequency.

From relations (\ref{ds:eq8}) and (\ref{ds:eq9}) it follows in particular that in limiting cases when $z \rightarrow 0$ the quantity $\varphi \rightarrow \varphi_{e}=\sqrt{\omega / 2 K_{e}}$, $\alpha \rightarrow \alpha_{e}=\sqrt{\omega / 2 K_{e}}$ and when $z \rightarrow \infty$, the quantity $\varphi \rightarrow \varphi_{f}=\sqrt{\omega / 2 K_{f}}, \alpha \rightarrow \alpha_{f}=\sqrt{\omega / 2 K_{f}}$.

Hence, taking into account (\ref{ds:eq7}), we arrive to the following expression for solution (\ref{ds:eq5}) 
\begin{equation}\label{ds:eq10}
\tilde{u}(r, t)=A e^{i \omega t} \frac{\mathrm{K}_{1}[r(\varphi+i \alpha)]}{\mathrm{K}_{1}\left[r_{c}(\varphi+i \alpha)\right]},
\end{equation}
which satisfies the boundary conditions and allows one to identify corresponding solution of the problem (\ref{ds:eq3}) -- (\ref{ds:eq4}) in the real domain 
 $$
u(r, t)=\operatorname{Im} \tilde{u}(r, t) .
$$

For further studies, assessing the influence zone of pulsations of filtration microflows on colmatant region, we need an asymptotics of modified Bessel function  $K_{1}(s)$  for its argument greater than one \cite{ref19} (p.690, formula 14.127)
\begin{equation}\label{ds:eq11}
\mathrm{K}_{1}(s) \approx \sqrt{\frac{\pi}{2 s}} e^{-s}.
\end{equation}

Substituting (\ref{ds:eq11}) into expression (\ref{ds:eq10}), we lead to the following relation 
\begin{equation}\label{ds:eq12}
\tilde{u}(r, t)=A \sqrt{\frac{r_{c}}{r}} e^{i \omega t} e^{-\left(r-r_{c}\right)(\varphi+i \alpha)} .
\end{equation}
 
Then, evaluating the imaginary part of (\ref{ds:eq12}), we obtain the real-valued solution at large argument of the Bessel function   for nonequilibrium filtration 
\begin{equation}\label{ds:eq13}
u(r, t)=A \sqrt{\frac{r_{c}}{r}} \sin \left[\omega t-\left(r-r_{c}\right) \alpha\right] e^{-\left(r-r_{c}\right) \varphi} .
\end{equation}

Such solutions describe a pulsating standing damping wave, under the influence of which oscillating pulsating microflows are formed, which help to increase the permeability of the formation, washing out the pore channels [5,6].

It is worth to note that similar pulsating fluid oscillations were observed during experimental studies \cite{ref10} and in theoretical investigations of the movement of a viscous fluid in the medium's pores under the influence of harmonic wave action \cite{ref14,ref15}.
The size of zone of influence of wave fields is not large due to their significant attenuation, but such a size is sufficient to affect colmatant area of the NWZ, where the filtration fluid flow is most significantly suppressed. Therefore, an important objective is to estimate the attenuation of filtering pulsating oscillations in NWZ in a wide range of frequencies, depending on the parameter $\theta$ and parameter $\tau$.  

Next, based on the theoretical studies outlined above, let us estimate the NWZ size, which is under the influence of the wave action caused by the operation of the acoustic emitter mounted directly in a well.

As an estimate of the size of this area, we choose the distance  $r_{V}$, at which the attenuation of the initial disturbance reaches a predetermined value. In other words, attenuation (or the size of the influence zone) is conveniently characterized by the following quantity
$$
\Delta=\frac{\max _{t} u\left(t, r_{V}\right)}{\max _{t} u\left(t, r_{c}\right)}.
$$

Taking into account solution (\ref{ds:eq12}), we get   or in relative units  
\begin{equation}\label{ds:eq14}
\Delta=\sqrt{\frac{1}{\bar{r}_{V}}} e^{-\left(\bar{r}_{V}-1\right) r_{c} \varphi}.
\end{equation}

If we fix the value of $\Delta$, then the derived expression can be regarded as an algebraic equation with respect to the parameter $\bar{r}_{V}$. It is convenient to rewrite equation (\ref{ds:eq14}) in the following form
\begin{equation}\label{ds:eq15}
\Phi(y)=\ln y+2 y r_{C} \varphi-2 \ln M, 
\end{equation}
where   $y=\bar{r}_{V}>0$, $M=\Delta^{-1} e^{r_{c} \varphi}$.

Since $\frac{\partial \Phi}{\partial y}=\frac{1}{y}+2 r_{c} \varphi>0$, then the function $\Phi$   is monotonically increasing. Let us show that it can take values of different signs. To do this, we derive 
 $$
\Phi\left(M^{2}\right)=2 M^{2} r_{c} \varphi>0 .
$$
Next, let us estimate the value of the function at the point $\Phi\left(M^{-k}\right)$, where the exponent $k>0$ is chosen from the condition that $\Phi\left(M^{-k}\right)<0$.

Hence, let $M>1$, then
$$
\Phi\left(M^{-k}\right)=-k \ln M+2 M^{-k} r_{c} \varphi-2 \ln M .
$$

Choose $k$ in such a way that the term $2 M^{-k} r_{c} \varphi=\varepsilon$ is such that $\varepsilon-2 \ln M<0$. Then $k=\frac{\ln 2 r_{c} \varphi / \varepsilon}{\ln M}>0$, and also $0<\varepsilon<\ln M^{2}$.Therefore, taking $0<\varepsilon<\min \left\{2 r_{c} \varphi, \ln M^{2}\right\}$, we evaluate $k$ and, moreover, it is valid $\Phi\left(M^{-k}\right)<0$.

Thus, the function $\Phi$ is the continuous at the interval $y \in(0 ; \infty)$ and admits the values of different signs on the interval $\left\lfloor M^{-k} ; M^{2}\right]$. Then by the Intermediate Value Theorem, this function possesses a root on the specified interval, moreover, due to monotonic increasing this root is unique.

This root can be evaluated by the simple iteration method. To apply this method, equation (\ref{ds:eq15}) can be rearrange, for instance, in the form
\begin{equation}\label{ds:eq16}
y=\frac{2 \ln M-\ln y}{2 r_{c} \varphi} \equiv \psi(y), 
\end{equation}
for which the iteration scheme is as follows  $y_{n+1}=\psi\left(y_{n}\right)$. Let us note that for a simple iteration, a sufficient condition for convergence is the constraint 
 $\left|\frac{d \psi}{d y}\right|<1$. Since $\left|\frac{d \psi}{d y}\right|=\frac{1}{2 r_{c} \varphi}$, then on the interval $y \in\left[M^{-k} ; M^{2}\right]$ the derivative $\left|\frac{d \psi}{d y}\right|<\frac{M^{k}}{2 r_{c} \varphi}=\frac{1}{\varepsilon}<1$.
 
It is worth to note that at low frequencies it is possible to occur the case when $\min \left\{2 r_{c} \varphi, \ln M^{2}\right\}<1$ and then application of algorithm is impossible.

But the condition $\varepsilon>1$ can be eliminated if we choose another form of representation (\ref{ds:eq16}). In particular, let us
multiply equation (\ref{ds:eq15}) by a number $(-\lambda)$ (here $\lambda$ is positive), add $y$ to both parts and finally obtain
$$
y-\lambda\left(\ln y+2 y r_{c} \varphi-\ln M\right)=y
$$

The quantity $\lambda$ is chosen from the condition that
$$
\left|\frac{d\left(y-\lambda\left(\ln y+2 y r_{c} \varphi-\ln M\right)\right)}{d y}\right|=\left|1-\lambda\left(\frac{1}{y}+2 r_{c} \varphi\right)\right|<1 .
$$
Since $0<M^{-k}<y$, then $1-\lambda\left(\frac{1}{y}+2 r_{c} \varphi\right)<1$ is valid for all positive $\lambda$. Another part of inequality
$$
-1<1-\lambda\left(\frac{1}{y}+2 r_{c} \varphi\right) \Rightarrow \lambda\left(\frac{1}{y}+2 r_{c} \varphi\right)<2
$$
is valid if we take $\lambda=\frac{2}{M^{k}+2 r_{C} \varphi}$. Indeed,
$$
\lambda\left(\frac{1}{y}+2 r_{c} \varphi\right)<\frac{2}{M^{k}+2 r_{c} \varphi}\left(M^{k}+2 r_{c} \varphi\right)=2 .
$$

Thus, the iteration process

\begin{equation}\label{ds:eq17}
y_{n+1}=y_{n}-\frac{2}{M^{k}+2 r_{c} \varphi}\left(\ln y_{n}+2 y_{n} r_{c} \varphi-\ln M\right).
\end{equation}

is convergent for arbitrary initial data without auxiliary constraints for $\varepsilon$.
Similar studies can be performed also for the case $M<1$.

\begin{figure}[h]
\begin{center}
\includegraphics[totalheight=3in]{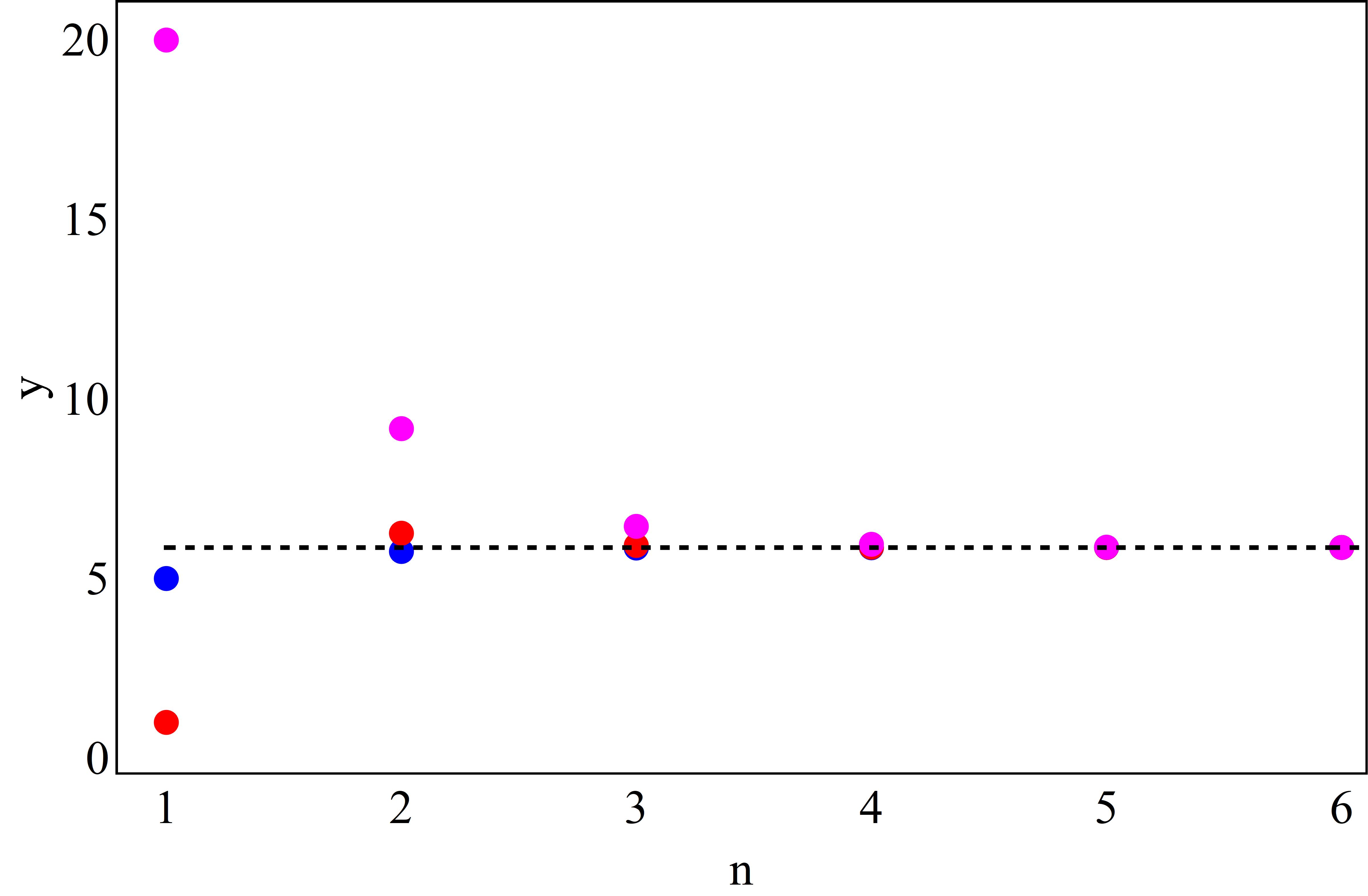}\hspace{0.5 cm} 
\end{center}
\caption{Iteration process for finding the solution of equation (\ref{ds:eq14}) by means of algorithm (\ref{ds:eq17}) at fixed parameter values  $\omega=100$ s$^{-1}$,  $\tau=0.03$s,  $ \theta=0.05$,  $K_{e}=2$ m/$s^{2}$, $r_{c}=0.1$ m. Dashed line corresponds to the root value evaluated numerically by the tools in-built in {\it Mathematica} package.
} \label{ds:fig01}
\end{figure}

To show the algorithm's work, let us fix the parameter values $\omega=100 \mathrm{~s}^{-1}$, $ \tau=0.03 \mathrm{~s}$, $\theta=0.05$, $ K_{e}=2 \mathrm{~m} / \mathrm{s}^{2}$, $r_{C}=0.1 \mathrm{~m}$, $\Delta=0.1$.
Then the quantity $M=13.385>1$. Since $\min \left\{2 r_{c} \varphi, \ln M^{2}\right\}=\min \{0.583,5.188\}=0.583<1$, then we can choose $\varepsilon$ for instance as $\varepsilon=0.5$ and apply algorithm (\ref{ds:eq17}). The process of algorithm convergence is shown in Fig.~\ref{ds:fig01} for different initial data. Analysis of Fig.~\ref{ds:fig01} indicates that the algorithm is convergent for a wide range of initial data (as it should be in accordance with the fulfillment of the sufficient convergence condition) with quite good rate of convergence (in fact, the third iteration provides the root).

Using the results obtained above, the relative influence radii $\bar{r}_{V}$ are calculated for different values of perturbation frequency $\omega$, time of relaxations $\tau$ and parameter $\theta$ when other parameters are fixed: $r_{c}=0.1 \mathrm{~m}, \mathrm{~K}_{e}=2 \mathrm{~m}^{2} / \mathrm{s}$, parameter $\Delta=0.1$.
 
 \begin{figure}[h]
\begin{center}
\includegraphics[totalheight=3in]{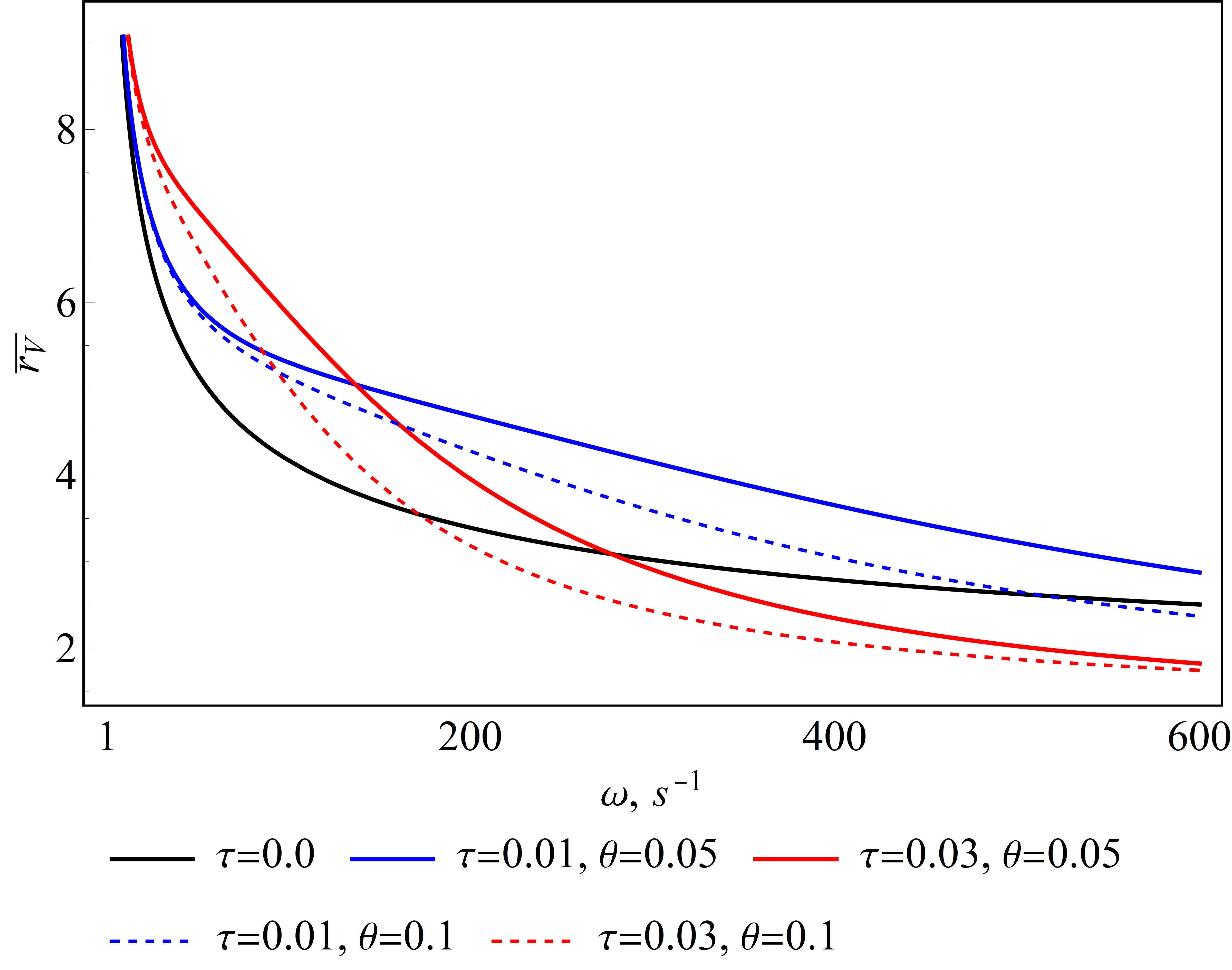}
\end{center}
\caption{The dependence of $\bar{r}_{V}$ on the frequency $\omega$ at fixed $\tau$ and $\theta$. The other parameters are $K_{e}=2 \mathrm{~m} / \mathrm{s}^{2}, r_{c}=0.1 \mathrm{~m}$.
} \label{ds:fig02}
\end{figure}

Thus, Fig.~\ref{ds:fig02} represents the solutions of equation (\ref{ds:eq14}) as a function $\bar{r}_{V}(\omega ; \tau, \theta)$ of frequency $\omega$, when $\tau$ and $\theta$ take discrete values, e.g. $\tau=0$ (absence of relaxing effects), $\tau=0.01, \quad \tau=0.03$, while $\theta=0.05$ and $\theta=0.1$ (dashed curves).

It is quite interesting the behavior of graphs at the edge points of domain of the function. When $\omega \rightarrow 0$ ( $\tau$ is fixed), the curves $\bar{r}_{V}$ approach the point $\bar{r}_{V}(0)=\Delta^{-2}$. When $\omega \rightarrow \infty$, then equation (\ref{ds:eq14}) reduces to the form
$$
\sqrt{\bar{r}_{V}}=\Delta^{-1} \exp \left(\left(1-\bar{r}_{V}\right) r_{c} \varphi\right),
$$
in which $\varphi \rightarrow \varphi_{f} \rightarrow \infty$ at $\omega \rightarrow \infty$. In this case the equation possesses the bounded solution if $\bar{r}_{V} \rightarrow 1$. Thus, we encounter the $0 \cdot \infty$ uncertainty, which can lead us to bounded result. Therefore, if solution (\ref{ds:eq14}) exists at $\omega \rightarrow \infty$, then this can be realized at $\bar{r}_{V}=1$ only, that is confirmed by the asymptotics of the graphs in Fig.~\ref{ds:fig02}. Hence, we can state that all curves $\bar{r}_{V}(\omega)$ possess two common points: $\omega=0$ and $\omega=\infty$.

No less interesting is the behavior of the graphs at varying $\tau$. Near $\omega=0$ for each curve $\bar{r}_{V}(\omega)$ there exists the interval (its size decreases when $\tau$ grows), on which the curve is close to equilibrium curve $\bar{r}_{V}(\omega; \tau=0)$. This indicates the weak influence of relaxation on the system's dynamics at low frequencies. Moreover, the amplitude of initial perturbation attenuates slowly such that its 10 -fold decrease is observed at distances of 6-8 well radii $r_{c}$. However, for higher $\omega$ the values $\bar{r}_{V}(\omega)$ attenuates quickly and for $\omega>200 \mathrm{~s}^{-1}$ the radius of influence zone is about $3 r_{c}$, i.e., high-frequency disturbances do not penetrate far into the formation, which is consistent with classical results.

As $\omega$ increases, relaxation plays a more prominent role, i.e. the more $\tau$, the more $\bar{r}_{V}$. However, due to the faster decline of the function $\bar{r}_{V}$, the intersection points of the curves corresponding to different $\tau$ appear. This means that only in a limited range of frequencies relaxation effects contribute to the growth of the influence zone, that is, the size of $\bar{r}_{V}$. For frequencies beyond this interval, relaxation behaves as an additional mechanism for dissipating the energy of oscillatory motion. As can be seen for the curve $\bar{r}_{V}(\omega ; \tau=0.03)$, at high frequencies the amplitude of the initial disturbance decreases by a factor of 10 already at distances of the order of $2 r_{c}$, which is less than for relaxation processes with shorter $\tau$ and under equilibrium conditions.

\begin{figure}[h]
\begin{center}
\includegraphics[totalheight=3in]{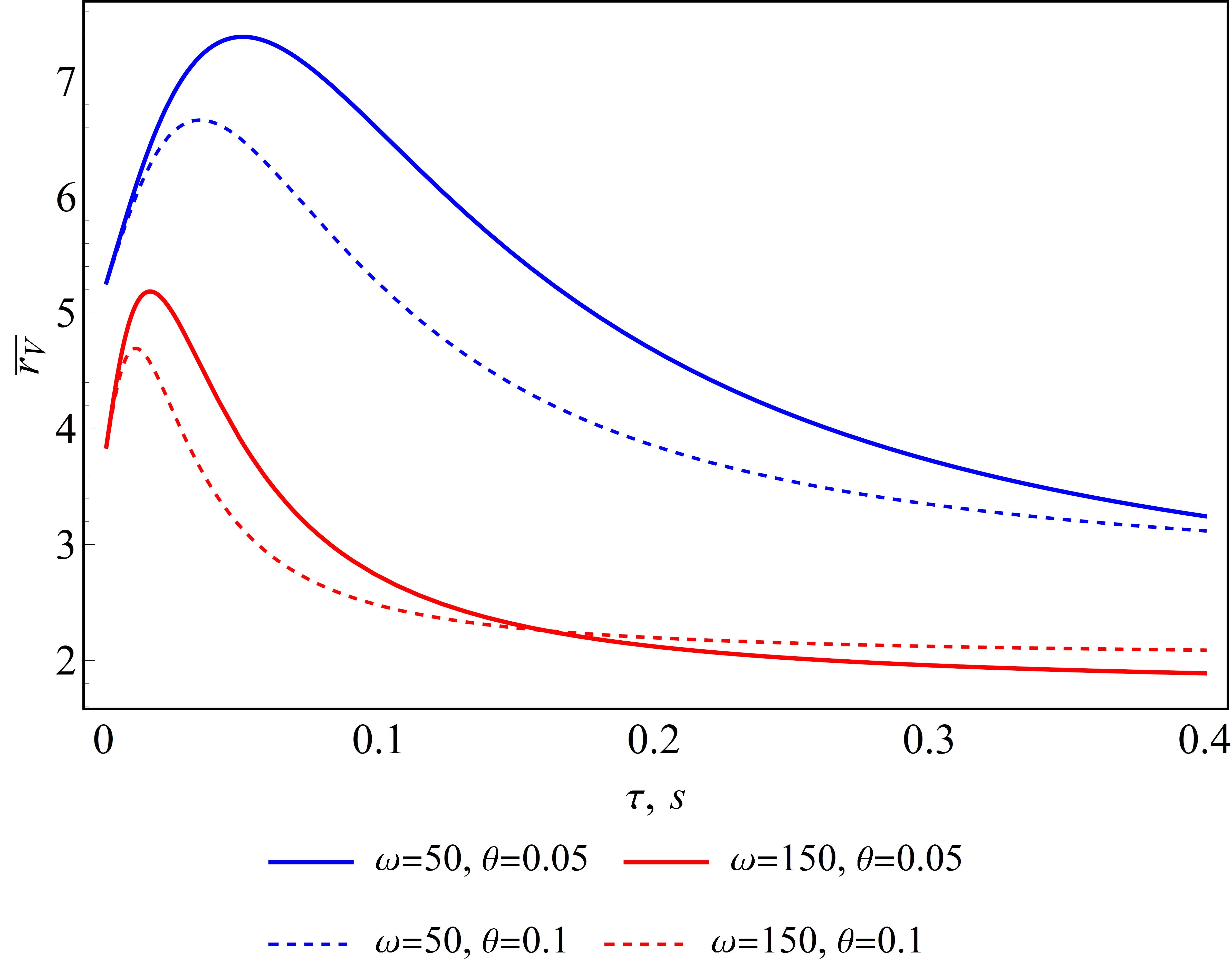}
\end{center}
\caption{The dependence $\bar{r}_{V}$ on time of relaxation $\tau$ at fixed $\omega$ and $\theta$.The parameters are as follows $K_{e}=2 \mathrm{~m} / \mathrm{s}^{2}, r_{c}=0.1 \mathrm{~m}$.
} \label{ds:fig03}
\end{figure}

Figure \ref{ds:fig03} exhibits the graphs of dependence $\bar{r}_{V}(\tau ; \omega, \theta)$ on the time of relaxation $\tau$ at fixed $\omega=50 \mathrm{~s}^{-1}$ and $\omega=150 \mathrm{~s}^{-1}$ (solid curves) when $\theta=0.05$.

In contrast to the graphs in Fig.~\ref{ds:fig02}, in intervals of small $\tau$, the curves $\bar{r}_{V}(\tau)$ in Fig.~\ref{ds:fig03} have local maxima on their profiles. To calculate their coordinates, we use the necessary extremum condition, applying it to equation (\ref{ds:eq14}) as an implicitly defined function. If we assume that $\bar{r}_{V}=y=y(\tau)$ and also according to
(\ref{ds:eq8}) $\varphi=\varphi(\tau)$, then the derivative is as follows
$$
\frac{d y}{d \tau}=2 r_{c} \frac{\Delta^{-1} \sqrt{y} e^{(1-y) r_{c} \varphi}-y^{2}}{1+2 y r_{c} \varphi} \cdot \frac{d \varphi}{d \tau}
$$

It is obvious that at the critical point $\tau=\tau_{c r}$ we get $\frac{d y}{d \tau}=0$, when $\frac{d \varphi}{d \tau}=0$. From this it follows the condition of existence of the local extreme value
\begin{equation}\label{ds:eq18}
\omega \tau_{c r}=\sqrt{\frac{3(1+\theta)-\sqrt{9+14 \theta+9 \theta^{2}}}{2 \theta^{2}}} . 
\end{equation}

Using expression (\ref{ds:eq18}), it is easy to show that at increasing $\theta$   the critical point shifts to the left.

If $\theta$ increases to 0.1, we get graphs $\bar{r}_{V}$ marked with dashed lines (Fig.\ref{ds:fig03}). Note that the curves corresponding to different $\theta$, intersect (compare the solid and dashed curves).

 \begin{figure}[h]
\begin{center}
\includegraphics[totalheight=3in]{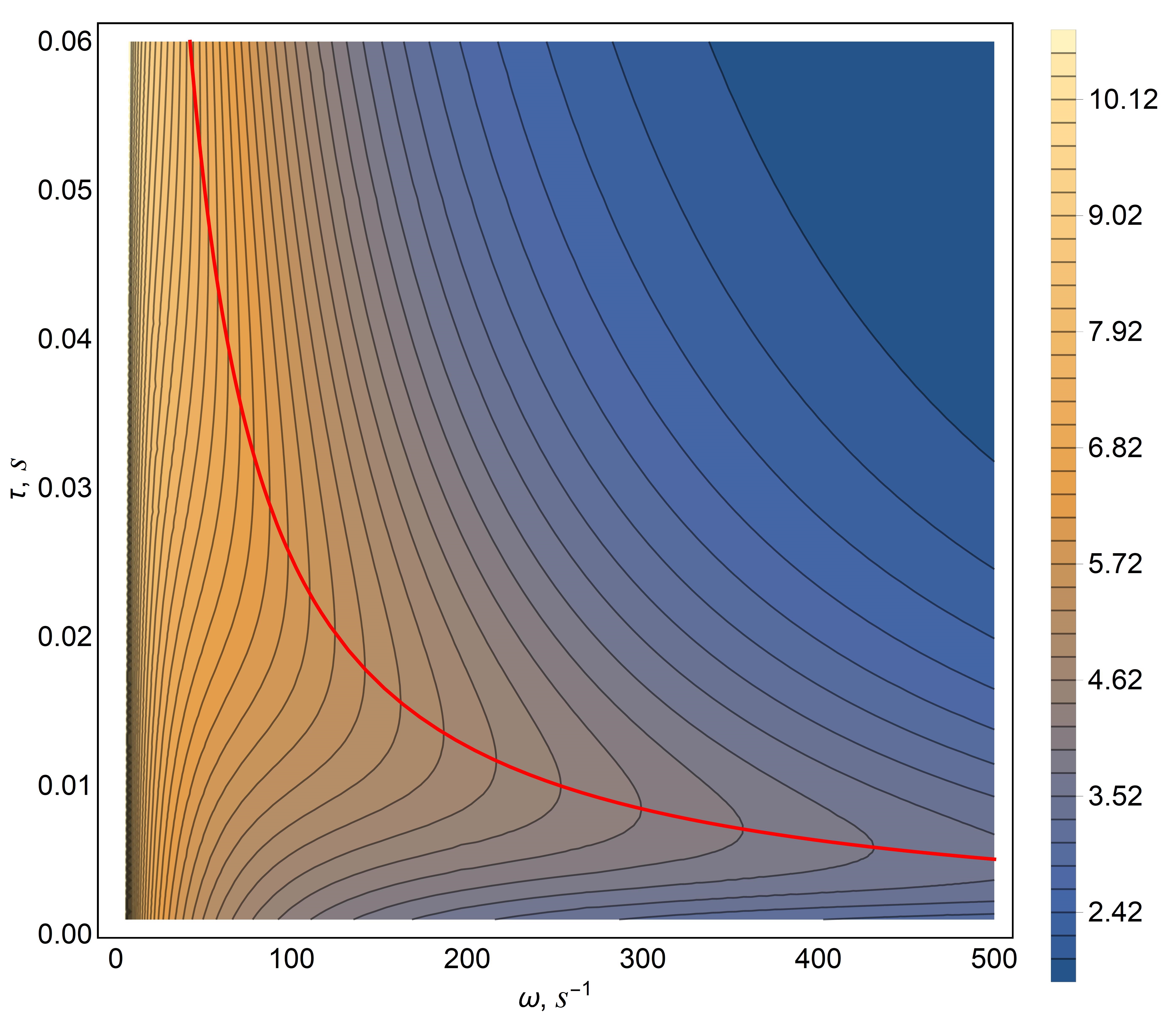}
\end{center}
\caption{Contour plot for the function $\bar r_V$  of  $\omega$   and $\tau$. The graph of function (\ref{ds:eq18}) is depicted by red solid curve. Here  $\theta=0.05$,  $K_{e}=2 \mathrm{~m} / \mathrm{s}^{2}$, $r_{c}=0.1 \mathrm{~m}$. 
} \label{ds:fig04}
\end{figure}

This means that increasing $\theta$ affects differently $\bar{r}_{V}$ on whole interval $\tau$. In the most interesting interval $\tau$, growth $\theta$ causes suppression of the maximum $\bar{r}_{V}$ and its shift to small values of $\tau$.

The mutual influence of frequency and relaxation time is clearly visible on the contour plot of Fig.~\ref{ds:fig04}. The red solid line marks the locus of local maxima, which are determined by condition (\ref{ds:eq18}) at $\theta=0.05$.

Thus, the selection of frequency $\omega$ for NWZ processing should be carried out in such a way as to stay close to the local maxima of the graph $\bar{r}_{V}(\tau; \omega, \theta)$. This, in turn, requires more detailed information about the reservoir and its saturating fluid.

\section{Concluding remark} 

In the conducted research, using the generalized dynamic Darcy law with one relaxation parameter, a mathematical model of the elastic regime of non-stationary non-equilibrium fluid filtration in a porous semi-bounded medium of a circular formation is considered.
The research deals with the boundary value problem of non-equilibrium filtering with harmonic perturbation at the boundary of a semi-confined reservoir and the additional condition that the solution is bounded at infinity. By the method of separation of variables, a non-stationary complex-valued solution is obtained in the form of a product of a harmonic function of time and a modified Bessel function of the second kind of the first order with respect to the spatial coordinate. Based on this solution, an asymptotic solution of the problem for large values of the argument of the Bessel function is constructed. This solution determines the pulsating fluid motion in a porous formation.
Using the derived solution, the damping of pulsations during non-equilibrium filtering is analyzed depending on the frequency of the wave action, the ratio of the piezoconductivity coefficients  $K_{f} / K_{e}$, and relaxation parameter. 

Graphs of the dependences of the size of the influence zone on the model parameters are plotted and the choice of parameters for optimal influence on the NWZ is analyzed. It is established that the sizes of the zones affected by vibration during non-equilibrium filtration exceed the sizes of these zones during equilibrium filtration processes.
A mathematical model of non-equilibrium fluid filtration is proposed and the results of theoretical studies are obtained, which are relevant for the development of wave technologies for the intensification of mineral resource extraction [20].
In the future, studies of non-equilibrium filtration in the case of a larger number of relaxation processes, as well as in the case of the dependence of formation permeability coefficients on the spatial coordinate, are of scientific and practical interest.

\section*{Acknowledgments}
The authors express their gratitude to D. Sci. Mykulyak Sergiy and D. Sci. Skurativskyi Sergii for useful discussions of the research findings. 

The research is partly supported by the  National Academy of Science of Ukraine, Project 0123U100181 and Project 0123U100183.

\end{document}